# IONORT: a Windows software tool to calculate the HF ray tracing in the ionosphere

A. Azzarone, C. Bianchi, M. Pezzopane*, M. Pietrella, C. Scotto, A. Settimi

*Istituto Nazionale di Geofisica e Vulcanologia, Via di Vigna Murata, 605, 00143 Rome, Italy*

michael.pezzopane@ingv.it

Tel.: +390651860525

Fax: +390651860397

**Abstract**

This paper describes an applicative software tool, named IONORT (IONOspheric Ray Tracing), for calculating a three-dimensional ray tracing of high frequency waves in the ionospheric medium. This tool runs under Windows operating systems and its friendly graphical user interface facilitates both the numerical data input/output and the two/three-dimensional visualization of the ray path. In order to calculate the coordinates of the ray and the three components of the wave vector along the path as dependent variables, the core of the program solves a system of six first order differential equations, the group path being the independent variable of integration. IONORT uses a three-dimensional electron density specification of the ionosphere, as well as by geomagnetic field and neutral particles-electrons collision frequency models having validity in the area of interest.

Keywords: ray tracing, ray path, ionospheric models.

## 1. Introduction

Ray tracing (RT) is a numerical technique used to determine the path of a high frequency (HF) radio wave in anisotropic and inhomogeneous media different from the vacuum (Budden, 1988). The technique works properly if the refractive index is assumed to be known in each point of the considered region. In the limits of the ray theory it is possible to approximate the wavelength to zero, simplifying consequently the differential equations describing the propagation of the wave in a suitable way, i.e. the ray path. Hence, three-dimensional (3-D) RT algorithms calculate the coordinates reached by the wave vector and its three components, the group time delay of the wave along the path and other optional quantities (geometrical and phase path, absorption, polarization, etc.). In order to accomplish these tasks, the RT programs integrate at least six differential equations, plus other equations when additional quantities, like for instance Doppler frequency shift, are required. These kinds of algorithms were developed in the 1950's (Haselgrove, 1955; Duziak, 1961; Croft and Gregory, 1963) for old mainframes that were able to give only a numerical output. Nowadays, these programs have been optimized and adapted to Over The Horizon radar applications (Coleman, 1998; Nickish, 2008) by using powerful computers and devices for a real-time use.

This paper deals with a software tool, named IONORT, whose RT algorithm is based on a system of first order differential equations with Hamiltonian formalism that are solved for a geocentric spherical coordinate system. The corresponding software (that can be downloaded at the site ftp://ftp.ingv.it/pub/adriano.azzarone/ionort_0.7.2.zip) is written in MATLAB for the input and the output routines, while the integration algorithm is derived from the one that was coded in Fortran by Jones and Stephenson (1974). In the next future, a whole package coded in MATLAB is planned. The ionosphere considered by this software tool is represented by 3-D ionospheric regional models elaborated at the Istituto Nazionale di Geofisica e Vulcanologia. An analytical standard Chapman modeled ionosphere (Chapman, 1931) useful mainly for test purpose complete the whole package.

## 2. Generality on the ionospheric ray tracing algorithm

Ray tracing techniques rely on a comprehensive specification of the ionosphere in terms of electron density, neutral particles-electrons collision frequency, and geomagnetic field. As an example, Fig. 1 shows a 3-D matrix of the electron density where in each cell $C_{ijk}$ (here, the indexes $i$, $j$, and $k$ are respectively the longitude, the latitude, and the altitude and these depend on the corresponding matrix resolution) the electron density of the ionospheric medium, and hence the corresponding complex refractive index $n$, has a defined value.

Fig. 1 shows also a possible ray path from a transmitting (TX) point to a receiving (RX) point across the cells of the 3-D electron density matrix with the frequency and direction of the wave, and the position of the TX point given as input. The RT algorithm integrates the following partial differential equations

$$\frac{d r_i}{d \tau} = \frac{\partial H(r_i, k_i)}{\partial k_i} \qquad (1.1)$$

$$\frac{d k_i}{d \tau} = \frac{\partial H(r_i, k_i)}{\partial r_i} \qquad (1.2)$$

where $i=1,..,4$, $H(r_i, k_i)$ is the Hamiltonian, $r_i$, and $k_i$, are respectively the generalized coordinates and momenta, while the independent variable $\tau$ must be a monotonic increasing quantity (represented in our case by the group path) (Weinberg, 1962). (1.1) and (1.2) are solved for a geocentric spherical coordinate system ($r$, $\theta$, $\varphi$), and according to the wave vector components, as shown in Fig. 2 (Bianchi and Bianchi, 2009; Bianchi et al., 2010). The essential differential equations that are integrated are given more explicitly in the Appendix A.

## 3. IONORT: description of the program

IONORT is structured in three main blocks:

a) INPUT GRAPHICAL USER INTERFACE;
b) INTEGRATION ALGORITHM;
c) OUTPUT GRAPHICAL USER INTERFACE.

Fig. 3 shows the flowchart of the IONORT application.

The block a), developed in MATLAB, firstly reads a file named "DATA_default.ini" to initialize the default inputs related to all the computational parameters needed by the ray-tracing algorithm. After this phase of initialization, it visualizes a graphical user interface (GUI, see Fig. 4 or Fig. 5) by which the user can modify the default inputs, and then it generates a file "DATA_in.txt" representing the user input for the integration executable code, written in Fortran, that is the block b). "DATA_in.txt" is then nothing but a copy of the file "DATA_default.ini" modified according to the choices made by the user. "DATA_in.txt" is then the actual input of the Fortran core, which reads it as a vector $W$ of 400 components. Table 1 shows the first 25 components of such a vector, in particular: the geographical coordinates of the TX point, the height (in km) of the TX and the RX points at which the program must start and stop respectively, the azimuth and elevation angles (in degrees), the wave operating frequency (in MHz), the polarization of the ray (ordinary or extraordinary), the geographical coordinates of the geomagnetic pole, the number of hops, and some needed constants like for instance the Earth radius. The user can modify some of these input parameters, as well as the analytical or numerical models representing the ionosphere (according to what is shown by Table 2), by filling and checking the corresponding boxes of the "Main parameters", "Step", "Model", and "Ray" frames of the GUI. Once the parameters have been set, the "RUN" button launches the integration algorithm.

The block b), which is the core of the application, is represented by this integration algorithm that is coded as a Fortran executable. In order to integrate step by step the differential equations illustrated in the Appendix A, this executable performs all the computational operations using either the 4-order Runge-Kutta (RK) method (Press et al., 1996) or the Adams-Bushford predictor and the Adams-Moulton corrector methods (ABAM) (Press et al., 1996). Using these, the ray path of the wave in spherical coordinates is calculated.

Once this task ended, the block c), besides saving the numerical output in a file "RToutput.txt", visualizes the results in the GUI where also 2-D and 3-D graphical elaborations of the ray path are performed. The 2-D visualization is plotted at the bottom of the GUI in a plane section having constant azimuth. The 3-D visualization is plotted on the right side of the GUI. The numerical output of some relevant parameters like the latitude and the longitude of the arrival point, the ground range distance on the Earth's surface, the maximum altitude of the path trajectory (apogee), and the time delay of the ray along the whole path (group delay) are shown in the "Results" frame of the GUI.

IONORT can run both with a fixed operating frequency and with a frequency-step procedure. The same is for the elevation and the azimuth angles. Fig. 4 and Fig. 5 show two examples of elaboration, by taking into account a TX point at 43.06°N of latitude and at 10.03°E of longitude, the former for a fixed frequency equal to 6 MHz and for a 5° elevation-step procedure from 0° to 30°, the latter for a fixed elevation angle equal to 15° and for a 2 MHz frequency-step procedure from 2 MHz to 24 MHz.

**4. Description of the integration computational code**

The main task of IONORT is the integration of the first-order differential equation system given in the Appendix A. The discrete form of the system can be write as

125 $$\frac{dy_i}{d\tau} = f_i(\tau, y_1, ...., y_N),\qquad(2)$$

127 where $i=1,...,$ N, $y_i$ are the dependent variables (coordinates and wave vector components) and $\tau$ is
128 the independent variable. For each of the six equations of the system, at the step $n+1$ the classical 4-
129 order RK formula gives

131 $$y_{n+1} = y_n + h\frac{1}{6}(k_1 + 2k_2 + 2k_3 + k_4),\qquad(3.1)$$

133 with

134 $$\tau_{n+1} = \tau_n + h,\qquad(3.2)$$

136 where $n$ and $h$ represent respectively the integration step number and the integration step value,
137 $k_1=f(\tau_n,y_n)$, $k_2=f(\tau_n+h/2,y_n+hk_1/2)$, $k_3=f(\tau_n+h/2,y_n+hk_2/2)$, and $k_4=f(\tau_n+h,y_n+hk_3)$. Alternatively, the
138 system can be solved using the ABAM method. In the RK method the independent variable $\tau$ can
139 assume values from tens of meters to a few kilometers, while in the ABAM method there is an
140 adaptive step according with a maximum and a minimum tolerated error.
141  With regard to the refractive index calculation routine, as we have already mentioned in the
142 paragraph 3, besides the possibility of considering an analytical standard Chapman modeled
143 ionosphere (Chapman, 1931), numerical representations of the ionosphere can also be considered.
144 The corresponding 3-D electron density matrixes (as the one shown in Fig. 1) are formed by cells
145 extending tens of kilometers in latitude and in longitude, and a few kilometers in altitude. At each
146 step of integration the algorithm evaluates the actual cell $C_{ijk}$ along the path, according to the
147 flowchart shown in Fig. 6, and hence takes the corresponding value of electron density, which is
148 considered the same inside the whole cell.

Concerning the numerical 3-D representation of the ionosphere, this is based either on the Adaptive Ionospheric Profiler (AIP) model developed by Scotto (2009) or on the model proposed by Pezzopane et al. (2011). Both models derive from the International Reference Ionosphere (Bilitza, 2008) and rely on the real-time autoscaling performed both by Autoscala (Pezzopane and Scotto, 2007, 2008, 2010) and by ARTIST (Reinisch and Huang, 1983, Galkin and Reinisch, 2008).

**5. Consistency check of the integration algorithm**

In order to check the error due to the integration algorithm, the time delay of the wave as it results from the RT computation performed by IONORT, and the time delay of the wave propagating along the oblique virtual path at the speed of light $c$, were compared at different frequencies. In order to calculate the latter time delay, a flat reflector is assumed at an altitude compatible with the vertical virtual height of reflection. The relation between the vertical and oblique frequencies is given by the secant law $f_v = f_{ob} \cos\varphi$ (Davies, 1990), where $\varphi$ is the incidence angle, $f_v$ is the vertical frequency, and $f_{ob}$ is the oblique frequency. The equality of the two time delays is assured by the Breit-Tuve and Martyn theorems (Davies, 1990) in case of a monotonically increasing electron density profile. Bianchi et al. (2011) ran such a test by employing a numerical electron density matrix, and the corresponding results are shown in Fig. 7. It came out that the IONORT ray tracing algorithm fits nearly perfectly the theory stated by the two aforementioned theorems. This means that the relative error $\Delta t_{error}$ between the time delay $t_{calc}$ calculated by IONORT and the simulated time delay $t_{virt}$, calculated according to the Breit-Tuve and Martyn theorems, is only due to the discrete integration step.

**6. Conclusions**

In this paper, an applicative software tool package running under Windows operating system, named IONORT, capable to solve the ray tracing for HF waves propagating in the ionosphere was described. The integration algorithm of IONORT is coded in Fortran, while the GUI managing the input needed to the integration algorithm, and the corresponding numerical and graphical output, is coded in MATLAB. This GUI facilitates noticeably the numerical input data entry made by the user and at the same time performs a useful 2-D/3-D visualization of the ray path.

From a numerical point of view, in order to calculate the coordinates of the ray and the three wave vector components along the path as dependent variables, IONORT solves at least six first order differential equations, the group time being the independent variable of integration.

The consistency of the integration algorithm was checked by comparing real and virtual time delays.

The possibility offered to the user of choosing among different ionospheric electron density models, having validity in the area of interest, gives IONORT the necessary flexibility. It is worth noting that this last feature makes IONORT a valuable tool to test the goodness of the 3-D electron density representation of the ionosphere calculated by a definite model. In fact, given a radio link for which oblique soundings are routinely carried out, IONORT gives the possibility to generate synthesized oblique ionograms over the same radio link. The comparison between synthesized and measured oblique ionograms, both in terms of the ionogram shape and in terms of the maximum usable frequency characterizing the radio path, offers a great opportunity to understand how well the model can represent the real conditions of the ionosphere (Angling and Khattatov, 2006). Anyhow, this issue will be presented and discussed in a forthcoming paper.

**Appendix A. Equation (1.1) and (1.2) in spherical coordinates**

In spherical coordinates the equations (1.1) and (1.2) become

$$\frac{\mathrm{d}r}{\mathrm{d}\tau} = \frac{\partial H}{\partial k_r}, \qquad (A.1)$$

199 $$\frac{d\theta}{d\tau} = \frac{1}{r}\frac{\partial H}{\partial k_\theta},\qquad\text{(A.2)}$$

200

201 $$\frac{d\varphi}{d\tau} = \frac{1}{r\sin\theta}\frac{\partial H}{\partial k_\varphi},\qquad\text{(A.3)}$$

202

203 $$\frac{dk_r}{d\tau} = -\frac{\partial H}{\partial r} + k_\theta\frac{d\theta}{d\tau} + k_\varphi\sin\theta\frac{d\varphi}{d\tau},\qquad\text{(A.4)}$$

204

205 $$\frac{dk_\theta}{d\tau} = -\frac{1}{r}\left(-\frac{\partial H}{\partial\theta} - k_\theta\frac{dr}{d\tau} + k_\varphi r\cos\theta\frac{d\varphi}{d\tau}\right),\qquad\text{(A.5)}$$

206

207 $$\frac{dk_\varphi}{d\tau} = -\frac{1}{r\sin\theta}\left(-\frac{\partial H}{\partial\varphi} - k_\varphi\sin\theta\frac{dr}{d\tau} - k_\varphi r\cos\theta\frac{d\theta}{d\tau}\right),\qquad\text{(A.6)}$$

208

209 where $H$ is the Hamiltonian, $k_r$, $k_\theta$, $k_\varphi$ (see Fig. 2) are the components of the wave vector along $r$, $\theta$,

210 and $\varphi$. The Hamiltonian $H$ is a constant during the ray propagation, and for the IONORT algorithm

211 the following relation was chosen

212

213 $$H(r,\theta,\varphi,k_r,k_\theta,k_\varphi) = \frac{1}{2}\mathrm{Re}\left[\frac{c^2}{\omega^2}\left(k_r^2 + k_\theta^2 + k_\varphi^2\right) - n^2\right],\qquad\text{(A.7)}$$

214

215 where $n$ is the phase refractive index, $c$ is the speed of light, and $\omega$ is the fixed angular frequency of

216 the wave.

217

**Fig. 1.** 3-D matrix of the electron density, where *i* and *j* vary with the latitude and longitude, and *k* with the altitude, respectively. Column on the left composed by *k* cells represents the vertical electron density profile $V_{ij}$. Ray path from a TX point to a RX point is represented by a dotted curve. Because of the involved distance, the 3-D matrix has a spherical shell shape.

**Fig. 2.** Geocentric reference system in spherical ($r, \theta, \varphi$) and Cartesian (*x, y, z*) coordinates. **k** is the wave vector and in red, in blue and in violet the corresponding projections along versors $i_r$, $i_\theta$, and $i_\varphi$.

**Fig. 3.** Flowchart of IONORT application.

**Fig. 4.** GUI of IONORT program. "Main parameters" and "Step" frames are related to the input data. "Model" frame shows the analytical and numerical ionospheric models that can be chosen by the user. "Ray" frame gives the user the possibility to choose between the two different polarization of the wave, ordinary or extraordinary. "Results" frame shows the numerical output values. "RUN" button launches the integration algorithm. "Reset" button clears all the different outputs. At the bottom and on the right side, the 2-D and the 3-D visualizations of the ray path are respectively shown by considering a TX point at 43.06°N of latitude and at 10.03°E of longitude, for a fixed frequency equal to 6 MHz and for a 5° elevation-step procedure from 0° to 30°.

**Fig. 5.** Same as Fig. 3 for a fixed elevation angle equal to 15° and for a 2 MHz frequency-step procedure from 2 MHz to 24 MHz.

**Fig. 6.** Flowchart of subroutine "cellfind".

325   **Fig. 7.** Time group delays, $t_{virt}$ and $t_{calc}$, calculated by employing a numerical electron density
326   matrix, and corresponding percentage relative error.

327

328   **Table 1.** First 25 components of the input vector *W*.

329

330   **Table 2.** Analytical and numerical electron density models that can be used by IONORT. NF and
331   WF stand for no magnetic field and with magnetic field respectively. Because of the ray path does
332   not change significantly at the employed operating frequencies, the models do not include the
333   neutral particles-electrons collision frequency.

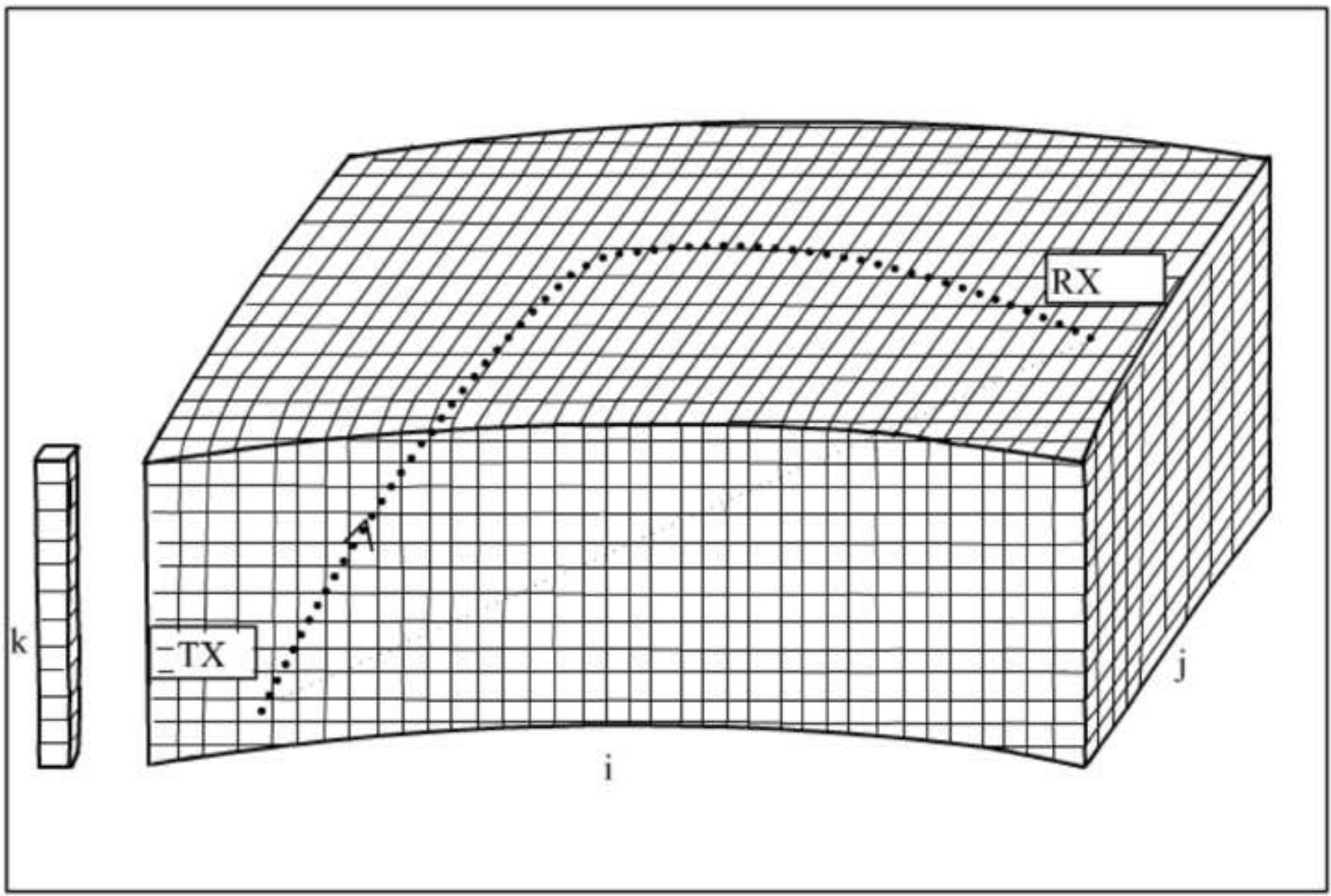

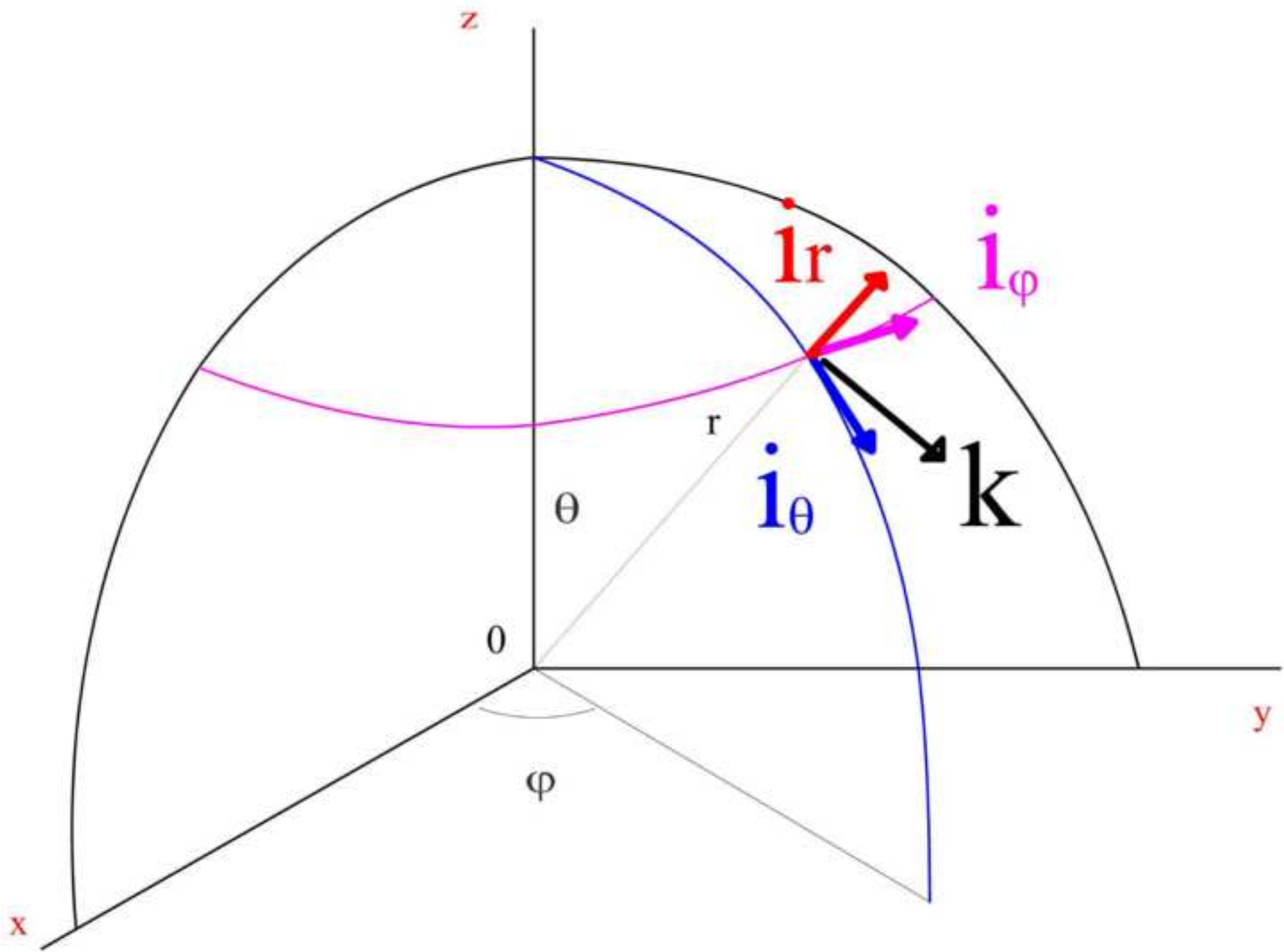

**Figure 3**
Click here to download high resolution image

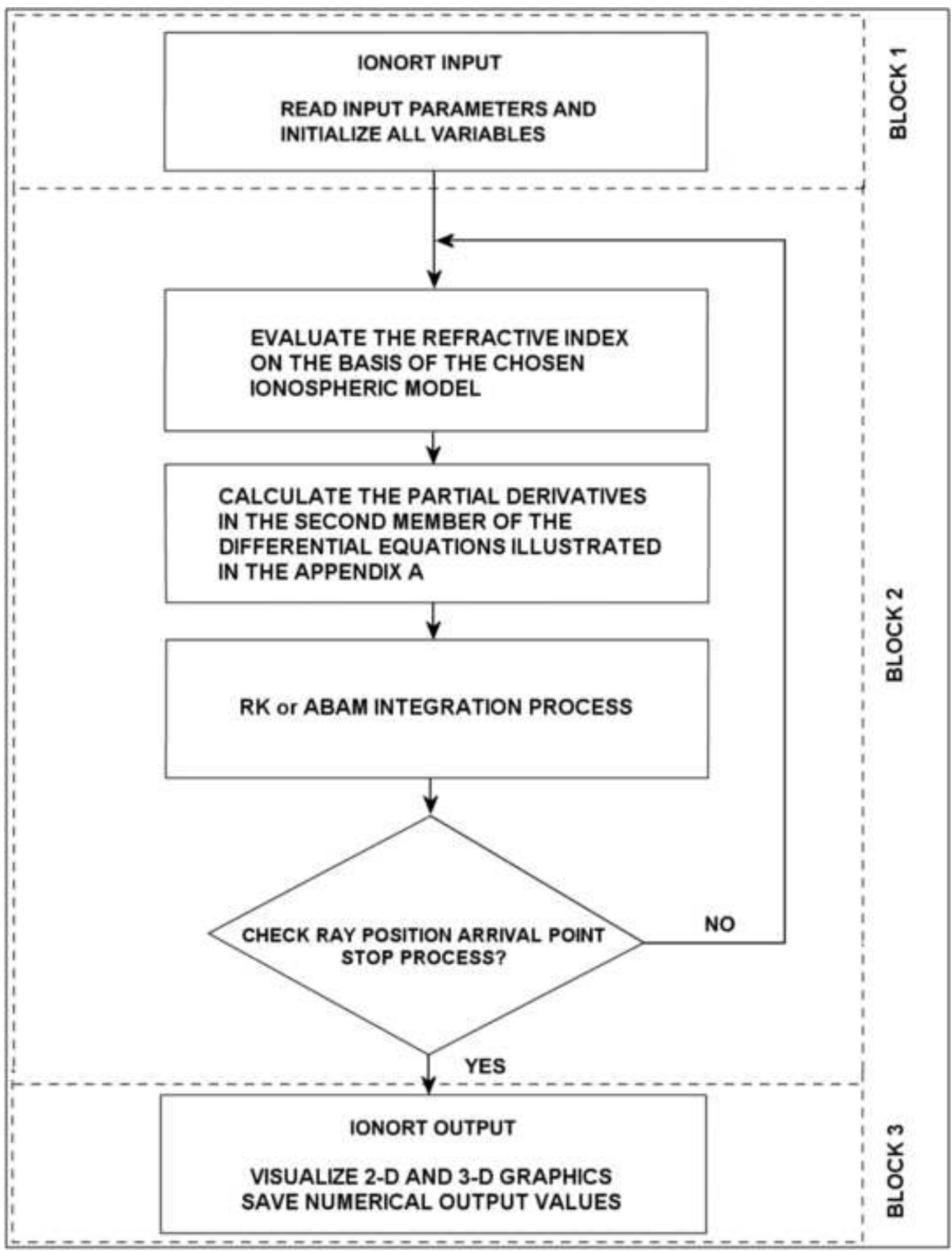

**Figure 4**
**Click here to download high resolution image**

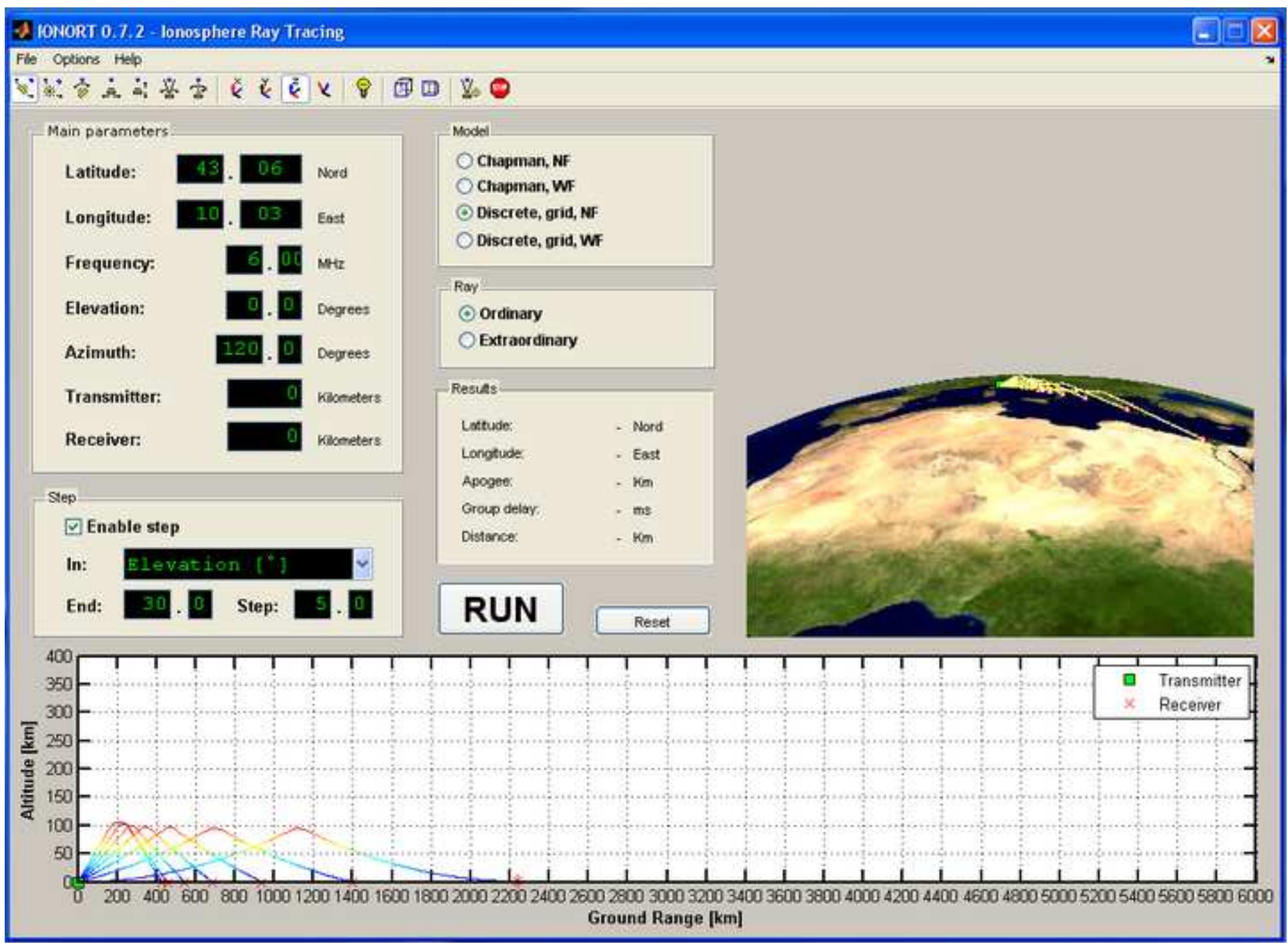

**Figure 5**
**Click here to download high resolution image**

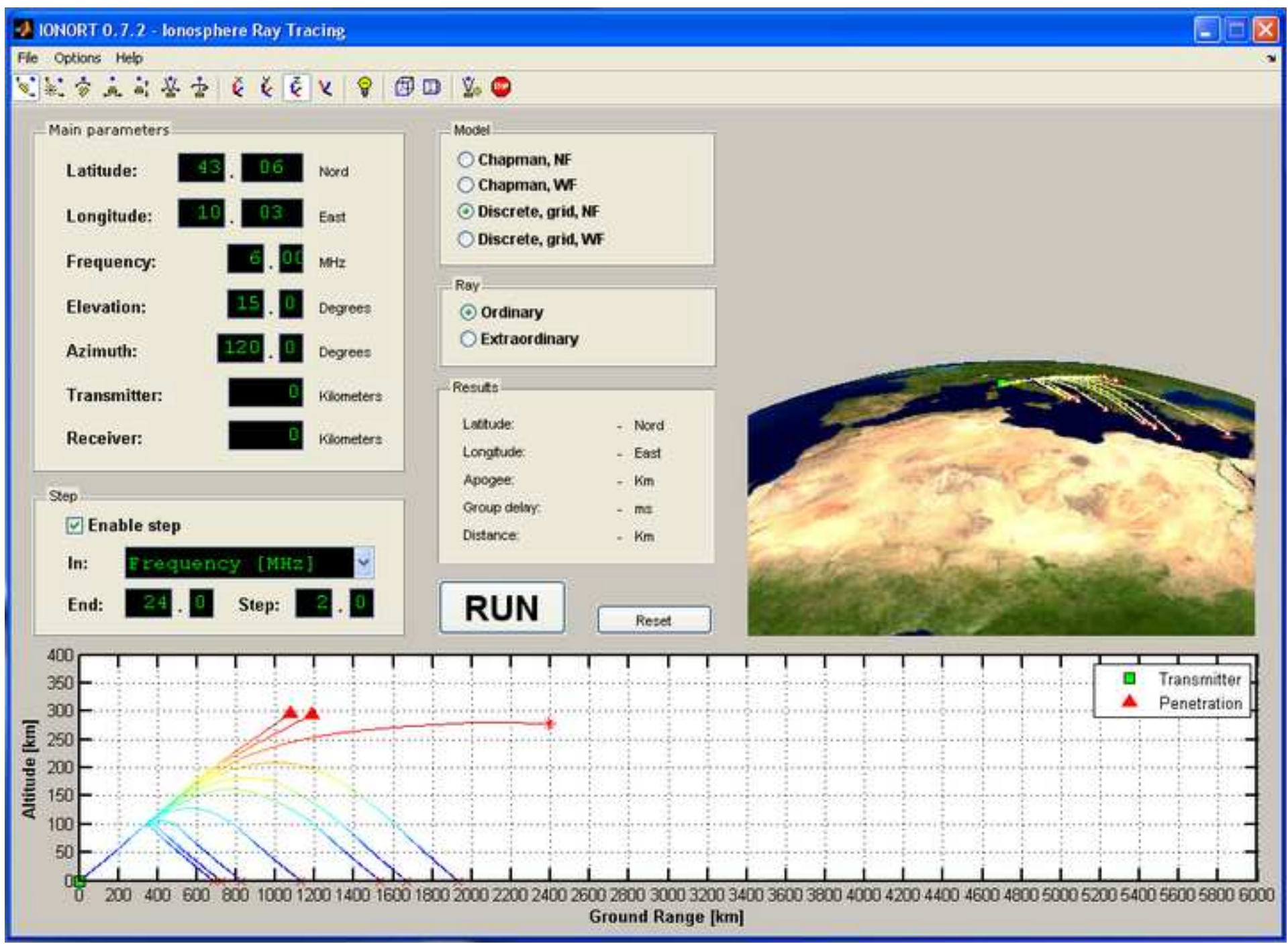

**Figure 6**
Click here to download high resolution image

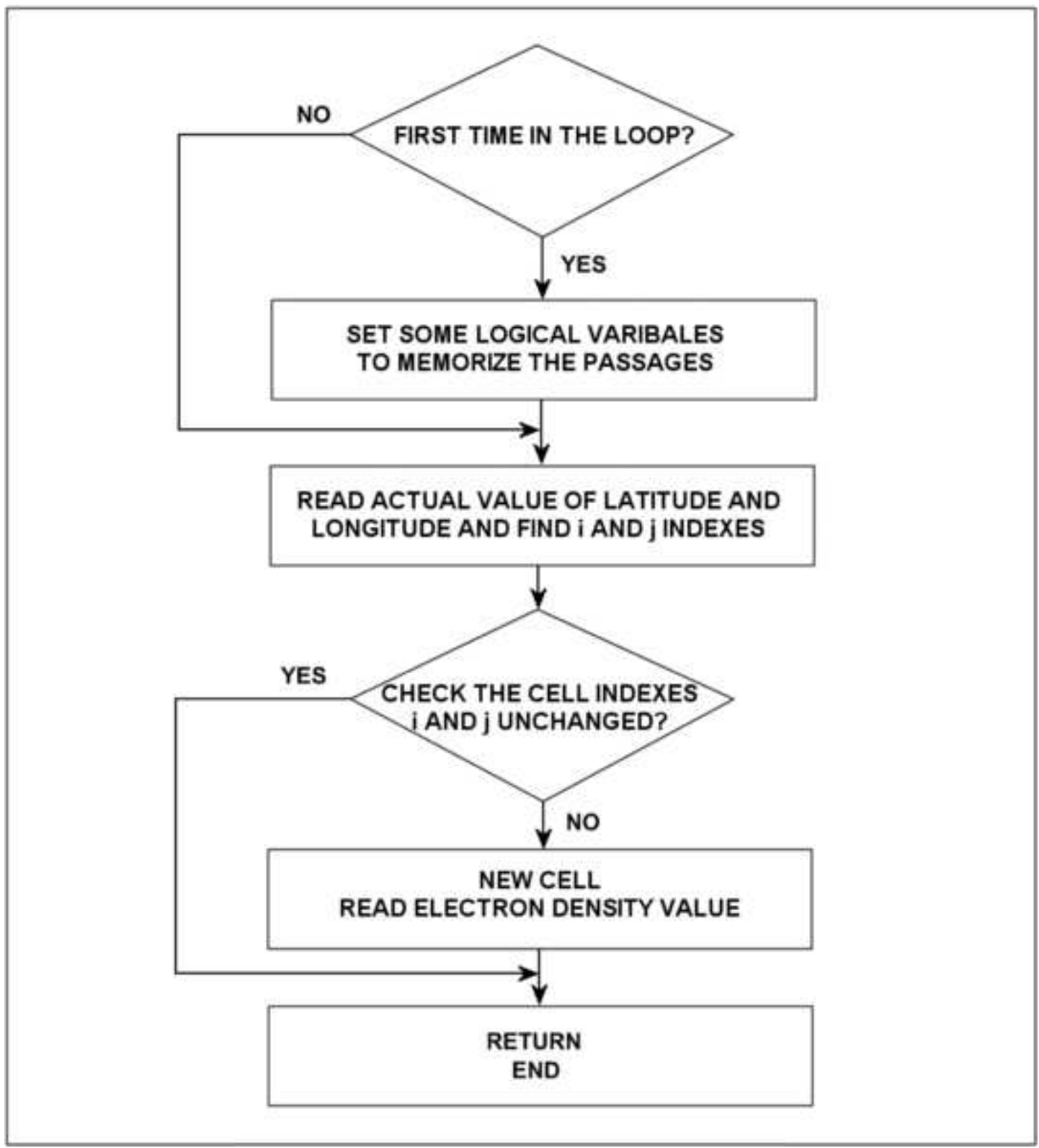

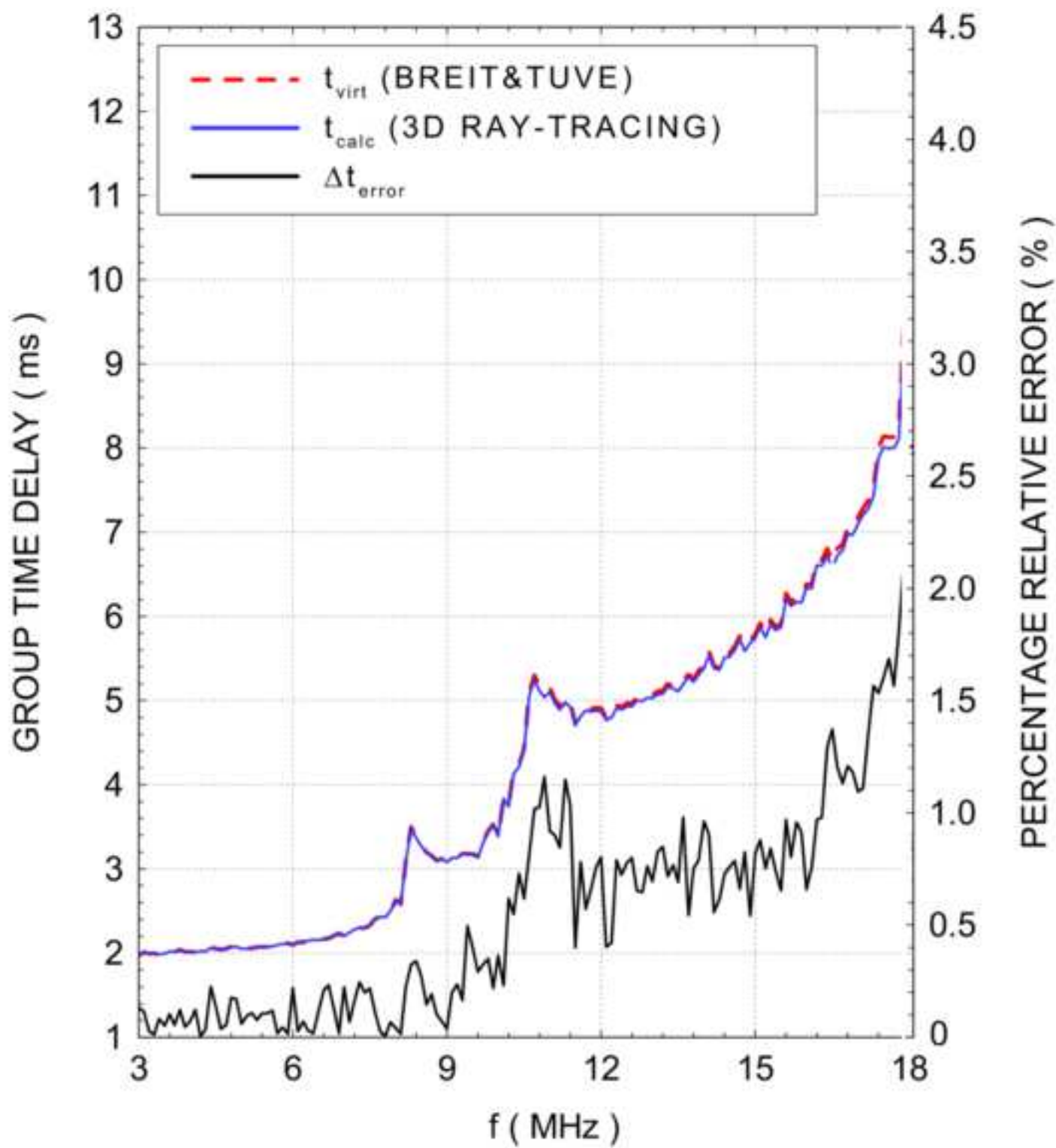

**Table 1**

| W vector component | Parameter | Description | Value |
| --- | --- | --- | --- |
| W1 | RAY | Radio Wave Mode | 1 = Ordinary<br>-1 = Extraordinary |
| W2 | EARTH | Earth Radius | 6371 km |
| W3 | XMTRH | Height of TX | km |
| W4 | TLAT | North geographic latitude of TX | rad |
| W5 | TLON | East geographic latitude of TX | rad |
| W6 | F | Frequency | MHz |
| W7 | FBEG | Initial frequency | MHz |
| W8 | FEND | Final frequency | MHz |
| W9 | FSTEP | Frequency step | MHz |
| W10 | AZI | Azimuth angle of transmission | rad |
| W11 | AZBEG | Initial azimuth | rad |
| W12 | AZEND | Final azimuth | rad |
| W13 | AZSTEP | Azimuth step | rad |
| W14 | BETA | Elevation angle of transmission | rad |
| W15 | ELBEG | Initial elevation | rad |
| W16 | ELEND | Final elevation | rad |
| W17 | ELSTEP | Elevation step | rad |
| W20 | RCVRH | Height of RX | km |
| W21 | ONLY | Reflected/not reflected rays | 0 = only reflected rays<br>1 = reflected/penetrating rays |
| W22 | HOP | Maximum number of hops | real |
| W23 | MAXSTP | Maximum number of steps for hops | real |
| W24 | PLAT | North geographic latitude of north geomagnetic pole | rad |
| W25 | PLON | East geographic longitude of north geomagnetic pole | rad |

**Table 2**

| Model | Description | Magnetic Field |
|---|---|---|
| **Chapman, NF** | Analytical electron density profiles | No |
| **Chapman, WF** | Analytical electron density profiles | Yes |
| **Discrete, grid, NF** | Numerical gridded electron density profiles | No |
| **Discrete, grid, WF** | Numerical gridded electron density profiles | Yes |

**Computer Code**

[Click here to download Computer Code: Computer Code.doc](Computer Code.doc)